\documentclass[12pt]{article}
\usepackage{amsmath,amssymb}
\usepackage{bm}
\usepackage{graphicx,color}
\usepackage{wrapfig}
\begin{document}
\title{
\begin{flushright}
\ \\*[-80pt] 
\begin{minipage}{0.2\linewidth}
\normalsize
\end{minipage}
\end{flushright}
{\Large \bf SUSY contributions to CP violations in 
$b\to s$ and $b\to d$ transitions facing on new data
\\*[20pt]}}

\author{
\centerline{Yusuke~Shimizu$^{1,}$\footnote{E-mail address: shimizu@muse.sc.niigata-u.ac.jp}, \ \ 
Morimitsu~Tanimoto$^{1,}$\footnote{E-mail address: tanimoto@muse.sc.niigata-u.ac.jp}, \ \  and \ \
Kei~Yamamoto$^{2,}$\footnote{E-mail address: yamamoto@muse.sc.niigata-u.ac.jp}}
\\*[20pt]
\centerline{
\begin{minipage}{\linewidth}
\begin{center}
$^1${\it \normalsize
Department of Physics, Niigata University,~Niigata 950-2181, Japan }
\\*[4pt]
$^2${\it \normalsize
Graduate~School~of~Science~and~Technology,~Niigata~University, \\ 
Niigata~950-2181,~Japan }
\end{center}
\end{minipage}}
\\*[70pt]}

\date{
\centerline{\small \bf Abstract}
\begin{minipage}{0.9\linewidth}
\vskip  1 cm
\small
We study the contribution of the gluino-squark mediated flavor changing process 
for the CP violation in $b\to s$ and $b\to d$ transitions 
facing on recent experimental data. 
The mass insertion parameters of squarks are constrained 
by the branching ratios of $b\to s\gamma $ and $b\to d\gamma $ decays. 
In addition, the time dependent CP asymmetries of 
$B^0\to \phi K_S$ and $B^0\to \eta ' K^0$ decays severely 
restrict the allowed region of the mass insertion parameter for the $b\to s$ transition. 
By using these constraints with squark and gluino masses of 
$1.5$~TeV, we predict the CP asymmetries of 
$B_s\to \phi \phi $, $B_s\to \eta '\phi $, and $B^0\to K^0\bar K^0$ decays, 
as well as the CP asymmetries in $b\to s\gamma $ and $b\to d\gamma $ decays. 
The CP violation in the $B_s\to \phi \phi$ decay is 
expected to be large owing to the squark flavor mixing, which 
will be tested at LHCb soon.
\end{minipage}
}

\begin{titlepage}
\maketitle
\thispagestyle{empty}
\end{titlepage}

\section{Introduction}
\label{sec:Intro}

The LHC experiments are now going on to discover new physics, 
for which the supersymmetry (SUSY) is one of the most attractive candidates. 
The SUSY signals have not yet observed 
although the Higgs-like events are almost confirmed~\cite{Higgs}. 
The lower bounds of the superparticle masses increase gradually. 
The squark and the gluino masses are expected 
to be larger than $1$ TeV~\cite{squarkmass}. 

On the other hand, the LHCb collaboration has reported 
new data of the CP violation of $B$ mesons and the branching ratios 
of rare $B$ decays~\cite{Bediaga:2012py,Aaij:2012ct,Lambert:2012gf}. 
The new physics is also expected to be indirectly 
found in the $B$ meson decays. 
For a long years the CP violation in the $K$ and $B^0$ mesons 
has been successfully understood within the framework of the standard model (SM), 
so called Kobayashi-Maskawa (KM) model~\cite{Kobayashi:1973fv}, 
where the source of the CP violation is the KM phase 
in the quark sector with three families. 
While, there are new sources of the CP violation if the SM is 
extended to the SUSY models. The soft squark mass matrices contain 
the CP-violating phases, which contribute to the flavor changing 
neutral current (FCNC) with the CP violation. 
Therefore, we expect the effect of the SUSY contribution in the CP-violating phenomena. 
However, the clear deviation from the prediction of the SM 
has not been observed yet in the LHCb experiment~\cite{Bediaga:2012py,Aaij:2012ct,Lambert:2012gf}. 
  
In our previous works~\cite{Hayakawa:2012ua,Shimizu:2012ru}, 
we studied the SUSY contribution, which comes from the gluino-squark mediated 
flavor changing process~\cite{King:2010np}-\cite{Ishimori:2011nv}. 
We used the only experimental data of the $b\to s\gamma $ decay to constrain 
the mass insertion (MI) parameters of squarks. And then, we predicted 
the CP violations of a few $b\to s$ transition processes. 
In the present paper, we give the systematic studies for the effect of 
the gluino-squark mediated flavor changing process in the CP violation of the 
$b\to s$ and $b\to d$ transitions. In order to obtain more precise numerical results, 
we take account of the QCD corrections for the SUSY contribution. 
Moreover, in order to constrain the MI parameters, we also input the 
recent experimental data, the time dependent CP asymmetries of $B^0$ non-leptonic decays, 
in addition to the experimental data of the $b\to s\gamma $ decay. 

The LHCb collaboration reported the time dependent CP asymmetry $S_{J/\psi \phi}$ 
in the non-leptonic $B_s$ decay, which gives a constraint of 
the SUSY contribution on the $b\to s$ transition. 
The CP asymmetry of $B_s \to \phi \phi $ 
is expected to be observed in the near future at LHCb~\cite{Lambert:2012gf}. 
If there is the contribution of the squark flavor mixing in the FCNC, we expect to observe 
the sizeable time dependent CP asymmetry in this process, 
in which the SM prediction is very small. 
   
The typical process of the $b\to s$ transition is the $b\to s\gamma $ decay, 
in which the experimental data of the branching ratio, 
the direct CP violation,  and the time dependent CP asymmetry 
$S_{K^*\gamma }$ have been reported. 
The SUSY contribution is also constrained by the data of the time dependent CP asymmetries 
in $B^0\to \phi K_S$ and $B^0\to \eta ' K^0$ decays~\cite{PDG,Amhis:2012bh}. 
 
On the other hand, the $b\to d$ transition also 
becomes available to investigate the SUSY contribution quantitatively 
taking account of the recent experimental branching ratio of the $b\to d\gamma $ 
decay~\cite{delAmoSanchez:2010ae,Crivellin:2011ba}. 
In this transition, the time dependent CP asymmetry of 
the $B^0 \to K^0 \bar K^0$ decay is an attractive one to search for the SUSY effect 
because the penguin amplitude dominates this process.
We also predict the time dependent CP asymmetry of $B^0 \to \rho \gamma $, $S_{\rho \gamma }$.   
   
The dominant contribution of the  SUSY is 
the gluino-squark mediated flavor changing process 
for the $B$ meson decays discussing in this work.
We present the constraint for the MI parameters
$(\delta _d^{LR})_{23}$ and $(\delta _d^{LR})_{13}$ by putting the experimental data,
where  squarks and the gluino  masses are at the TeV scale.
By using these MI parameters, we predict the
CP violation of the $B$ meson decays, in which the interesting  one is  
the $B_s\to \phi \phi $ decay. The CP violation in this decay will be measured 
 at LHCb in the near future. 

In section 2, we present the formulation of the  gluino-squark contribution
on the CP violation of $B$ mesons in our framework. 
In section 3, we discuss the $b\to s$ transition, and present numerical 
predictions for the direct CP violation and time dependent CP asymmetries 
in $B^0\to K^*\gamma $, $B_s\to \phi \phi $, and $B_s\to \eta' \phi$ decays. 
In section 4, we discuss the $b\to d$ transition, and present numerical 
predictions for the direct CP violation and time dependent CP asymmetries 
in $B^0\to \rho \gamma $ and $B^0\to K^0\bar K^0$ decays. 
Section 5 is devoted to the summary.


\section{Squark flavor mixing in CP violation of $B$ mesons}
\label{sec:Deviation}

Let us present the framework of the calculations for the contribution 
of the squark flavor mixing, 
which is the coupling among down-type quarks, down-type squarks, and the gluino. 
The effective Hamiltonian for the $\Delta B=1$ process is given as 
\begin{equation}
H_{eff}=\frac{4G_F}{\sqrt{2}}\left [\sum _{q'=u,c}V_{q'b}V_{q'q}^*
\sum _{i=1,2}C_iO_i^{(q')}-V_{tb}V_{tq}^*
\sum _{i=3-6,7\gamma ,8G}\left (C_iO_i+\widetilde C_i\widetilde O_i\right )\right ],
\end{equation}
where $q=s,d$. The local operators are given as 
\begin{align}
&O_1^{(q')}=(\bar q_\alpha\gamma _\mu P_Lq_\beta')
(\bar q_\beta'\gamma ^\mu P_Lb_\alpha),
\qquad O_2^{(q')}=(\bar q_\alpha\gamma _\mu P_Lq_\alpha')
(\bar q_\beta'\gamma ^\mu P_Lb_\beta), \nonumber \\
&O_3=(\bar q_\alpha\gamma _\mu P_Lb_\alpha)\sum _Q(\bar Q_\beta\gamma ^\mu P_LQ_\beta),
\quad O_4=(\bar q_\alpha\gamma _\mu P_Lb_\beta)\sum _Q(\bar Q_\beta\gamma ^\mu P_LQ_\alpha), \nonumber \\
&O_5=(\bar q_\alpha\gamma _\mu P_Lb_\alpha)\sum _Q(\bar Q_\beta\gamma ^\mu P_RQ_\beta),
\quad O_6=(\bar q_\alpha\gamma _\mu P_Lb_\beta)\sum _Q(\bar Q_\beta\gamma ^\mu P_RQ_\alpha), \nonumber \\
&O_{7\gamma }=\frac{e}{16\pi ^2}m_b\bar q_\alpha\sigma ^{\mu \nu }P_Rb_\alpha
F_{\mu \nu }, 
\qquad O_{8G}=\frac{g_s}{16\pi ^2}m_b\bar q_\alpha\sigma ^{\mu \nu }
P_RT_{\alpha\beta}^ab_\beta G_{\mu \nu }^a,
\end{align}
where $P_R=(1+\gamma _5)/2$, $P_L=(1-\gamma _5)/2$, and $\alpha $, $\beta $ are color 
indices, and $Q$ is taken to be $u,d,s,c$ quarks. 
Here, $C_i$'s and $\widetilde C_i$'s are the Wilson coefficients, 
and $\widetilde O_i$'s are the operators by replacing $L(R)$ with $R(L)$ 
in $O_i$. In this paper, $C_i$ includes both SM contribution and gluino one, 
such as $C_i=C_i^{\rm SM}+C_i^{\tilde g}$, where 
$C_i^{\text{SM}}$'s are given in Ref.~\cite{Buchalla:1995vs}. 

In order to estimate the SUSY contribution for $C_i^{\tilde g}$, 
we take the most popular ansatz, 
a degenerate SUSY breaking mass spectrum for the flavor structure of squarks. 
In the super-CKM basis, we can parametrize 
the soft scalar masses squared of the down-type squarks, 
$M^2_{\tilde d_{LL}}$, $M^2_{\tilde d_{RR}}$, 
$M^2_{\tilde d_{LR}}$, and $M^2_{\tilde d_{RL}}$ 
as follows: 
\begin{align}
M^2_{\tilde d_{LL}}&=m_{\tilde q}^2
\begin{pmatrix}
1+(\delta _d^{LL})_{11} & (\delta _d^{LL})_{12} & (\delta _d^{LL})_{13} \\
(\delta _d^{LL})_{12}^* & 1+(\delta _d^{LL})_{22} & (\delta _d^{LL})_{23} \\
(\delta _d^{LL})_{13}^* & (\delta _d^{LL})_{23}^* & 1+(\delta _d^{LL})_{33}
\end{pmatrix}, \nonumber \\
M^2_{\tilde d_{RR}}&=m_{\tilde q}^2
\begin{pmatrix}
1+(\delta _d^{RR})_{11} & (\delta _d^{RR})_{12} & (\delta _d^{RR})_{13} \\
(\delta _d^{RR})_{12}^* & 1+(\delta _d^{RR})_{22} & (\delta _d^{RR})_{23} \\
(\delta _d^{RR})_{13}^* & (\delta _d^{RR})_{23}^* & 1+(\delta _d^{RR})_{33}
\end{pmatrix}, \nonumber \\
M^2_{\tilde d_{LR}}&=(M_{\tilde d_{RL}}^2)^\dagger =m_{\tilde q}^2
\begin{pmatrix}
(\delta _d^{LR})_{11} & (\delta _d^{LR})_{12} & (\delta _d^{LR})_{13} \\
(\delta _d^{LR})_{21} & (\delta _d^{LR})_{22} & (\delta _d^{LR})_{23} \\
(\delta _d^{LR})_{31} & (\delta _d^{LR})_{32} & (\delta _d^{LR})_{33}
\end{pmatrix},
\end{align}
where $m_{\tilde q}$ is the average squark mass, and 
$(\delta _d^{LL})_{ij}$, $(\delta _d^{RR})_{ij}$, $(\delta _d^{LR})_{ij}$, 
and $(\delta _d^{RL})_{ij}$ are called as 
the mass insertion (MI) parameters. 

The Wilson coefficients of the gluino contribution 
$C_i^{\tilde g}$ are given as follows~\cite{Endo:2004fx}: 
\begin{align}
C_3^{\tilde g}(m_{\tilde g})&\simeq \frac{\sqrt{2}\alpha _s^2}{4G_FV_{tb}V_{tq}^*m_{\tilde q}^2}(\delta _d^{LL})_{k3}
\left [-\frac{1}{9}B_1(x)-\frac{5}{9}B_2(x)-\frac{1}{18}P_1(x)-\frac{1}{2}P_2(x)\right ], \nonumber \\
C_4^{\tilde g}(m_{\tilde g})&\simeq \frac{\sqrt{2}\alpha _s^2}{4G_FV_{tb}V_{tq}^*m_{\tilde q}^2}(\delta _d^{LL})_{k3}
\left [-\frac{7}{3}B_1(x)+\frac{1}{3}B_2(x)+\frac{1}{6}P_1(x)+\frac{3}{2}P_2(x)\right ], \nonumber \\
C_5^{\tilde g}(m_{\tilde g})&\simeq \frac{\sqrt{2}\alpha _s^2}{4G_FV_{tb}V_{tq}^*m_{\tilde q}^2}(\delta _d^{LL})_{k3}
\left [\frac{10}{9}B_1(x)+\frac{1}{18}B_2(x)-\frac{1}{18}P_1(x)-\frac{1}{2}P_2(x)\right ], \nonumber \\
C_6^{\tilde g}(m_{\tilde g})&\simeq \frac{\sqrt{2}\alpha _s^2}{4G_FV_{tb}V_{tq}^*m_{\tilde q}^2}(\delta _d^{LL})_{k3}
\left [-\frac{2}{3}B_1(x)+\frac{7}{6}B_2(x)+\frac{1}{6}P_1(x)+\frac{3}{2}P_2(x)\right ], \nonumber \\
C_{7\gamma }^{\tilde g}(m_{\tilde g})&\simeq -\frac{\sqrt{2}\alpha _s\pi }{6G_FV_{tb}V_{tq}^*m_{\tilde q}^2}
\Bigg [(\delta _d^{LL})_{k3}\left (\frac{8}{3}M_3(x)-
\mu \tan \beta \frac{m_{\tilde g}}{m_{\tilde q}^2}\frac{8}{3}M_a(x)\right )
+(\delta _d^{LR})_{k3}\frac{m_{\tilde g}}{m_b}\frac{8}{3}M_1(x)\Bigg ], \nonumber \\
C_{8G}^{\tilde g}(m_{\tilde g})&\simeq -\frac{\sqrt{2}\alpha _s\pi }{2G_FV_{tb}V_{tq}^*m_{\tilde q}^2}
\Bigg [(\delta _d^{LL})_{k3}\Bigg \{ \left (\frac{1}{3}M_3(x)+3M_4(x)\right ) \nonumber \\
&-\mu \tan \beta \frac{m_{\tilde g}}{m_{\tilde q}^2}\left (\frac{1}{3}M_a(x)+3M_b(x)\right )\Bigg \} 
+(\delta _d^{LR})_{k3}\frac{m_{\tilde g}}{m_b}\left (\frac{1}{3}M_1(x)+3M_2(x)\right )\Bigg ],
\label{Coeff}
\end{align}
where $k=2, 1$ correspond to $b\to q$ ($q=s,d$) transitions, respectively. 
Here the double mass insertion is included in $C_{7\gamma }^{\tilde g}(m_{\tilde g})$ 
and $C_{8G}^{\tilde g}(m_{\tilde g})$. 
The Wilson coefficients $\widetilde C_i^{\tilde g}(m_{\tilde g})$'s are 
obtained by replacing $L(R)$ with $R(L)$ in $C_i^{\tilde g}(m_{\tilde g})$'s. 
The loop functions in Eq.(\ref{Coeff})  are presented 
in our previous work~\cite{Hayakawa:2012ua}. 
In our calculations, $C_{7\gamma}$ and $C_{8G}$ give dominant contributions 
to the CP violations in $b\to s$ and $b\to d$ transitions. 
The effective Wilson coefficients of $C_{7\gamma}(m_b)$ and 
$C_{8G}(m_b)$ are given at the leading order of QCD as follows~\cite{Buchalla:1995vs}: 
\begin{equation}
\begin{split}
C_{7\gamma}^{\tilde g}(m_b)
&= \zeta C_{7\gamma}^{\tilde g}(m_{\tilde g})
+\frac{8}{3}(\eta-\zeta) C_{8G}^{\tilde g}(m_{\tilde g}), \cr
C_{8G}^{\tilde g}(m_b)
&=\eta C_{8G}^{\tilde g}(m_{\tilde g}),
\end{split}
\end{equation}
where 
\begin{equation}
\zeta=\left ( 
 \frac{\alpha_s(m_{\tilde g})}{\alpha_s(m_t)} \right )^{\frac{16}{21}}
 \left ( 
 \frac{\alpha_s(m_t)}{\alpha_s(m_b)} \right )^{\frac{16}{23}} \ , \qquad
 \eta=\left ( 
 \frac{\alpha_s(m_{\tilde g})}{\alpha_s(m_t)} \right )^{\frac{14}{21}}
 \left ( 
 \frac{\alpha_s(m_t)}{\alpha_s(m_b)} \right )^{\frac{14}{23}} \ .
 \end{equation}

Let us discuss the time dependent CP asymmetries of $B^0$ and $B_s$ decaying 
into the final state $f$, which are defined as~\cite{Aushev:2010bq} 
\begin{equation}
S_f=\frac{2\text{Im}\lambda _{f}}{1+|\lambda_{f}|^2}\ , \qquad
C_f=\frac{1-|\lambda_{f}|^2}{1+|\lambda_{f}|^2}\ ,
\label{sf}
\end{equation}
where 
\begin{equation}
\lambda_{f}=\frac{q}{p} \bar \rho\ , \qquad 
\frac{q}{p}\simeq \sqrt{\frac{M_{12}^{q*}}{M_{12}^{q}}}, \qquad 
\bar \rho \equiv 
\frac{\bar A(\bar B_q^0\to f)}{A(B_q^0\to f)}.
\label{lambdaf}
\end{equation}
Here $M_{12}^q(q=s,d)$ are the dispersive parts of the
$B_q$-$\bar B_q$ mixing, in which 
the quark-squark-gluino interaction contributes in addition to the SM one.
 The MI parameters $(\delta _d^{LL})_{k3}$ and  $(\delta _d^{RR})_{k3}$
 $(k=2,1)$ are constrained by 
CP violations in the $\Delta B=2$ transition as discussed in our previous 
works~\cite{Hayakawa:2012ua,Shimizu:2012ru}.
 On the other hand, 
  $(\delta _d^{LR})_{k3}$ and  $(\delta _d^{RL})_{k3}$
 $(k=2,1)$ are constrained in the $\Delta B=1$ transition.

In the $B^0\to J/\psi  K_S$ and $B_s\to J/\psi  \phi$ decays,
 we write $\lambda_{J/\psi  K_S}$ and $\lambda_{J/\psi  \phi}$
in terms of phase factors, respectively:
\begin{equation}
\lambda_{J/\psi  K_S}\equiv 
-e^{-i\phi _d}, \qquad \lambda _{J/\psi \phi } \equiv e^{-i\phi _s}.
\label{new}
\end{equation}
The recent experimental data of these phases are \cite{Amhis:2012bh,ICHEP2012} 
\begin{equation}
\sin \phi _d=0.679\pm 0.020\ , \qquad \phi _s=-0.002\pm 0.083\pm 0.027 \ .
\label{phasedata}
\end{equation}
 We expect the SUSY contribution to be included in these observed values.

Since the $B^0\to J/\psi K_S$ process occurs at the tree level in  the SM, 
the CP asymmetry mainly originates from  $M_{12}^d$.
Although the $B^0\to \phi K_S$ and $B^0\to\eta 'K^0$ decays 
are penguin dominant ones,
their CP asymmetries also come from  $M_{12}^d$ in the SM.
Then, the CP asymmetries of
 $B^0\to J/\psi K_S$,  $B^0\to \phi K_S$, and 
$B^0\to \eta 'K^0$ decays are expected to be the same magnitude.

On the other hand, if the squark flavor mixing contributes to the decay 
at the one-loop level, its magnitude could be  comparable 
to the SM penguin one 
in  $B^0\to \phi K_S$ and $B^0\to \eta 'K^0$ decays, 
but the squark flavor mixing contribution is tiny in the $B^0\to J/\psi K_S$ 
decay because this process is at the tree level in the SM. 
Therefore, there is a possibility to 
find the SUSY contribution  by  observing 
the different CP asymmetries among those processes~\cite{Endo:2004dc}.
 
The time dependent CP asymmetry $S_{ J/\psi K_S}$ has been precisely measured. On the other hand, 
PDG~\cite{PDG} and Heavy Flavor Averaging Group (HFAG)~\cite{Amhis:2012bh} 
presented considerably different values for $S_{\phi K_S}$ while almost 
same one for $S_{\eta 'K^0}$. 
Each of the observed ones in HFAG is consistent with the SM prediction. 
In order to get conservative constraints, we take the data of 
these time dependent CP asymmetries in HFAG~\cite{Amhis:2012bh}, which are 
\begin{equation}
S_{ J/\psi K_S}=0.679\pm 0.020 \ , \qquad 
S_{\phi K_S}= 0.74^{+0.11}_{-0.13}\ , \qquad 
S_{\eta 'K^0}= 0.59\pm 0.07\ .
\label{Sfdata}
\end{equation}
These values may be regarded to be  same within the experimental error bar. 
Thus, the experimental values are consistent with the prediction of the SM. 
In other words, 
these data severely constrain the MI parameters 
$(\delta _d^{LR})_{23}$ and $(\delta _d^{RL})_{23}$ in our following analyses.


\section{The $b \to s $ transition}
\label{sec:bstransitions}

At first we discuss the contributions of the squark flavor mixing for the $b\to s$ transition, 
which are given in terms of the MI parameters $(\delta _d^{LL})_{23}$, 
$(\delta _d^{RR})_{23}$, $(\delta _d^{LR})_{23}$, and $(\delta _d^{RL})_{23}$. 
These MI parameters are constrained by the experimental data of $B$ meson decays. 


Let us show the formulation of the $b\to s$ transition.
The CP asymmetries $S_f$ of Eq.~(\ref{sf}) in the $b\to ss{\bar s}$ transition are 
one of the most important processes when we investigate the new physics. 
The CP asymmetries $S_f$ for $B^0\to \phi K_S$ and $B^0\to \eta 'K^0$ 
are given in terms of $\lambda_f$ in Eq.~(\ref{lambdaf}):
\begin{align}
\lambda _{\phi K_S,~\eta 'K^0}&=-e^{-i\phi _d}\frac{\displaystyle \sum _{i=3-6,7\gamma ,8G}
\left (C_i^\text{SM}\langle O_i \rangle +C_i^{\tilde g}\langle O_i \rangle +
\widetilde C_i^{\tilde g}\langle \widetilde O_i \rangle \right )}
{\displaystyle \sum _{i=3-6,7\gamma ,8G}\left (C_i^{\text{SM}*}\langle O_i \rangle 
+C_i^{{\tilde g}*}\langle O_i \rangle +\widetilde C_i^{{\tilde g}*}\langle \widetilde O_i 
\rangle \right )}~,
\label{asymBd}
\end{align}
where $\langle O_i \rangle $ is the abbreviation of $\langle f|O_i|B^0\rangle $. 
It is noticed $\langle \phi K_S|O_i|B^0\rangle =\langle \phi K_S|\widetilde O_i|B^0\rangle $ 
and $\langle \eta 'K^0|O_i|B^0\rangle =-\langle \eta 'K^0|\widetilde O_i|B^0\rangle $, 
because these final states have different parities~\cite{Endo:2004dc,Khalil:2003bi}. 
Since the dominant term comes from the gluon penguin $C_{8G}^{\tilde g}$, 
the decay amplitudes of $f=\phi K_S$ and $f=\eta 'K^0$ are given as follows: 
\begin{align}
\bar A(\bar B^0 \to \phi K_S)
& \propto C_{8G}(m_b) + {\tilde C}_{8G}(m_b), \nonumber \\
\bar A(\bar B^0 \to \eta '\bar K^0)
& \propto C_{8G}(m_b) - {\tilde C}_{8G}(m_b).
\end{align}
Since ${\tilde C}_{8G}(m_b)$ is suppressed compared to $C_{8G}(m_b)$ in the SM, 
the magnitudes of the time dependent CP asymmetries 
$S_f \ (f=J/\psi \phi, \ \phi K_S,\  \eta 'K^0)$ are almost same in the SM prediction. 
However, the squark flavor mixing gives the unsuppressed ${\tilde C}_{8G}(m_b)$, 
then, the CP asymmetries in those decays are expected to be deviated among them. 
Therefore, those experimental data  give us the tight constraint for $C_{8G}(m_b)$ and ${\tilde C_{8G}}(m_b)$. 

We have also $\lambda _{f}$ for $B_s\to \phi \phi $ and $B_s\to \phi \eta '$ as follow: 
\begin{align}
\lambda _{\phi \phi ,\phi \eta '}&=e^{-i\phi _s}\frac{\displaystyle \sum _{i=3-6,7\gamma ,8G}
C_i^\text{SM}\langle O_i \rangle +C_i^{\tilde g}\langle O_i \rangle +
\widetilde C_i^{\tilde g}\langle \widetilde O_i \rangle }
{\displaystyle \sum _{i=3-6,7\gamma ,8G}
C_i^{\text{SM}*}\langle O_i \rangle +C_i^{{\tilde g}*}
\langle O_i \rangle +\widetilde C_i^{{\tilde g}*}\langle \widetilde O_i \rangle }~,
\label{asymBs}
\end{align}
with $\langle \phi \phi |O_i|B_s\rangle =-\langle \phi \phi |\widetilde O_i|B_s\rangle $ 
and $\langle \phi \eta '|O_i|B_s\rangle =\langle \phi \eta '|\widetilde O_i|B_s\rangle $. 
The decay amplitudes of $f=\phi \phi $ and $f=\phi \eta '$ are given as follows: 
\begin{align}
\bar A(\bar B_s \to \phi \phi )
& \propto C_{8G}(m_b) - {\tilde C}_{8G}(m_b), \nonumber \\
\bar A(\bar B_s \to \phi \eta ')
& \propto C_{8G}(m_b) + {\tilde C}_{8G}(m_b).
\end{align}
Since  $C_{8G}\langle O_{8G}\rangle $ 
and $\tilde C_{8G}\langle \tilde O_{8G}\rangle $ 
dominate these amplitudes,  our numerical results are insensitive
 to the hadronic matrix elements.
In order to obtain precise results,
we also take account of the small contributions 
from other Wilson coefficients $C_i~(i=3,4,5,6)$ and $\tilde C_i~(i=3,4,5,6)$ 
in our calculations. 
We estimate each hadronic matrix element 
by using the factorization relations in Ref.~\cite{Harnik:2002vs}: 
\begin{equation*}
\langle O_{3} \rangle =\langle O_{4} \rangle =\left( 1+\frac{1}{N_c} \right) \langle O_{5} \rangle,
\quad \langle O_{6} \rangle =\frac{1}{N_c}\langle O_{5} \rangle,
\end{equation*}
\begin{equation}
\langle O_{8G} \rangle =\frac{\alpha _s(m_b)}{8 \pi }
\left( -\frac{2 m_b}{ \sqrt{\langle q^2 \rangle }}\right )
\left( \langle O_4 \rangle +\langle O_6 \rangle -\frac{1}{N_c}(\langle O_3 \rangle 
+\langle O_5 \rangle )\right ), 
\end{equation}
where $\langle q^2 \rangle ={\rm 6.3~GeV^2}$ 
and $N_c=3$ is the number of colors. 
One may worry about the reliability of these naive factorization relations. 
However, this approximation has been justified numerically 
in the relevant $b\to s$ transition as seen in the calculation of PQCD~\cite{Mishima:2003wm}. 

Let us discuss the contribution of the MI parameters 
to $C_{8G}^{\tilde g}$ in the Eq.~(\ref{Coeff}). 
Since the loop functions are of same order 
and $m_{\tilde q}\simeq m_{\tilde g}$, the ratio of the $LL$ component 
and the $LR$ one is $(\delta _d^{LL})_{23}\times \mu \tan \beta /m_{\tilde q}$ to 
$(\delta _d^{LR})_{23}\times m_{\tilde q}/m_b$. 
If ${\cal O}(\mu \tan \beta )\simeq {\cal O}(m_{\tilde q})$ and 
$m_{\tilde q}\gtrsim 1$ TeV, the $LR$ component may contribute significantly 
to $C_{8G}^{\tilde g}$ due to the enhancement factor 
$m_{\tilde q}/m_b={\cal O}(10^2)$. 
For example, in the case of $(\delta _d^{LL})_{23}=10^{-2}$ 
and $(\delta _d^{LR})_{23}=10^{-3}$, the $LR$ component 
dominates $C_{8G}^{\tilde g}$, while it is minor 
in $M_{12}^{d}$~\cite{Hayakawa:2012ua,Shimizu:2012ru}. 
Actually, the magnitude of $(\delta _d^{LL})_{23}$ is at most $10^{-2}$, 
which was estimated in our previous works~\cite{Hayakawa:2012ua,Shimizu:2012ru}. 
In our following calculations, we take $|(\delta _d^{LL})_{23}|\lesssim 10^{-2}$. 


We can also constrain the SUSY contribution from the $b\to s\gamma $ decay. 
Here we discuss three observable values, those are the branching ratio $\text{BR}(b\to s\gamma )$, 
the direct CP asymmetry $A_\text{CP}^{b\to s\gamma }$, 
and the time dependent CP asymmetry of $B^0 \to K^* \gamma $, $S_{K^* \gamma }$. 
The branching ratio BR$(b\to q\gamma )$($q=s,d$) is a typical process to investigate the new physics. 
It is given as~\cite{Buras:1998raa} 
\begin{equation}
\frac{\text{BR}(b\to q\gamma )}
{\text{BR}(b\to ce\bar {\nu _e})}
=
\frac{|V_{tq}^*V_{tb}|^2}
{|V_{cb}|^2}
\frac{6 \alpha }{\pi f(z)}
(|C_{7\gamma }(m_b)|^2+|{\tilde C}_{7\gamma }(m_b)|^2),
\label{Brbqgamma}
\end{equation}
where 
\begin{equation}
f(z)
=
1-8z+8z^3-z^4-12z^2 \text{ln}z~,\qquad 
z = \frac{m_{c,pole}^2}{m_{b,pole}^2}.
\end{equation}
Here $C_{7\gamma }(m_b)$ and $\tilde{C}_{7\gamma }(m_b)$ include both contributions 
from the SM and the gluino-squark mediated flavor changing process 
at the $m_b$ scale. 
As seen in Eq.~(\ref{Coeff}), MI parameters $(\delta _d^{LR})_{k3}$ dominate both 
$C_{7\gamma }^{\tilde g}$ and $C_{8G}^{\tilde g}$. 
Therefore, we should discuss the contribution 
from $(\delta _d^{LR})_{k3}$ in our numerical calculations. 


We can also estimate the direct CP violation $A_{\text{CP}}^{b\to q\gamma }$ 
in the $b\to q\gamma $ decay ($q=s,d$), which is given as~\cite{Kagan:1998bh} 
\begin{align}
A_{\text{CP}}^{b\to q\gamma } 
&=
\left .
\frac{\Gamma (\bar {B}\to X_q\gamma ) - \Gamma (B\to X_{\bar q} \gamma )}
{\Gamma (\bar B\to X_q\gamma ) + \Gamma (B\to X_{\bar q}\gamma )}
\right |_{E_{\gamma } > (1-\delta ) E_{\gamma }^{\text{max}}} \nonumber \\
&=
\frac{\alpha _s(m_b)}
{|C_{7\gamma }|^2+|{\tilde C}_{7\gamma }|^2}
\Bigg [
\frac{40}{81}
\text{Im}\small [C_2 C_{7\gamma }^*\small ]
-
\frac{8 z}{9}\small [v(z)+b(z,\delta )\small ]
\text{Im}\Big [\left (1+\frac{V_{uq}^* V_{ub}}{V_{tq}^* V_{tb}}\right )
C_2 C_{7\gamma }^*\Big ] \nonumber \\
&
-\frac{4}{9}
\text{Im}\small [ C_{8G} C_{7\gamma }^* + {\tilde C}_{8G} {\tilde C}_{7\gamma }^*\small ]
+
\frac{8z}{27}
b(z,\delta ) \text{Im}\Big [\left (1+\frac{V_{uq}^* V_{ub}}{V_{tq}^* V_{tb}} \right )
C_2 C_{8G}^*\Big ]\Bigg ],
\label{directbqgamma}
\end{align}
where $v(z)$ and $b(z,\delta )$ are explicitly given in~Ref.\cite{Kagan:1998bh}, 
and $C_i$, ${\tilde C}_i$ ($i=7\gamma,8G$) include both the SM and SUSY contributions at the $m_b$ scale. 


The time dependent CP asymmetry $S_{K^* \gamma}$ in the $B^0 \to K^*\gamma $ decay 
is also important measure of the CP violation:
\begin{equation}
S_{K^*\gamma }
=
\frac{2 {\rm Im}(-e^{-i\phi _d} {\tilde C}_{7\gamma}(m_b)/C_{7\gamma }(m_b))}
{|{\tilde C}_{7\gamma }(m_b)/C_{7\gamma }(m_b)|^2+1}.
\label{Kstargamma}
\end{equation}
This CP violation comes from the interference between $C_{7\gamma }(m_b)$ and 
${\tilde C}_{7\gamma }(m_b)$~\cite{Endo:2004fx,Atwood:1997zr}. 
In the  SM, 
${\tilde C}_{7\gamma }^\text{SM}(m_b)/C_{7\gamma }^\text{SM}(m_b)\propto m_s/m_b$ for this process.
Therefore, $S_{K^*\gamma }$ is suppressed~\cite{Atwood:1997zr}. 
However, $S_{K^*\gamma }$ could be enhanced owing to the squark flavor mixing. 

Our setup in our calculations are shown as follows. 
We take $\mu \tan \beta $ to be $1$~TeV, 
and set $|(\delta _d^{LL})_{23}|\simeq |(\delta _d^{RR})_{23}|\lesssim 10^{-2}$ 
following from our previous works~\cite{Hayakawa:2012ua,Shimizu:2012ru}. 
Then, the contribution of these MI parameters to $C_{7\gamma }^{\tilde g}$ and 
$C_{8G}^{\tilde g}$ are minor. 
On the other hand, $(\delta _d^{LR})_{23}$ and $(\delta _d^{RL})_{23}$
are severely constrained by magnitudes of $C_{7\gamma }$ and 
$C_{8G}$. 
In addition, we suppose $|(\delta _d^{LR})_{23}|=|(\delta _d^{RL})_{23}|$. 
Then, we can parametrize the MI parameters as follows: 
\begin{equation}
(\delta _d^{LR})_{23}=|(\delta _d^{LR})_{23}|e^{2i\theta _{23}^{LR}},
\qquad (\delta _d^{RL})_{23}=|(\delta _d^{LR})_{23}|e^{2i\theta _{23}^{RL}}.
\label{MILR}
\end{equation}

Now we show numerical analysis in our setup. 
In our following numerical calculations, 
we fix the squark mass and the gluino mass as 
\begin{equation}
m_{\tilde q}=1.5~\text{TeV},\qquad m_{\tilde g}=1.5~\text{TeV},
\label{SUSYmass}
\end{equation}
which are consistent with recent lower bound of these masses at LHC~\cite{squarkmass}. 

In our analysis, the present experimental data of $\text{BR}(b\to s\gamma )$, 
$S_{J/\psi K_S}$, $S_{\phi K_S}$, and $S_{\eta 'K^0}$ give tight constraints for MI parameters. 
Here we put the experimental data \cite{PDG} 
\begin{equation}
\text{BR}(b\to s\gamma )({\rm exp})=(3.53 \pm 0.24)\times 10^{-4}, 
\end{equation}
on the other hand, the SM has predicted  \cite{Misiak:2006zs}
\begin{equation}
\text{BR}(b\to s\gamma )({\rm SM})=(3.15 \pm 0.23)\times 10^{-4} .
\end{equation}
Therefore, there is a room for the contribution of the gluino-squark mediated 
flavor changing process.\footnote{In our analysis, we do not take account of 
the contribution of the charged Higgs and chargino in $b\to s \gamma $.} 
For $S_{J/\psi K_S}$, $S_{\phi K_S}$, and $S_{\eta 'K^0}$, 
we put the data in Eq.(\ref{Sfdata}). In the SM, these magnitudes of $S_f$ 
agree with among them. 
 
\begin{figure}[h!]
\begin{minipage}[]{0.45\linewidth}
\hspace{4cm}(a)

\includegraphics[width=7.5cm]{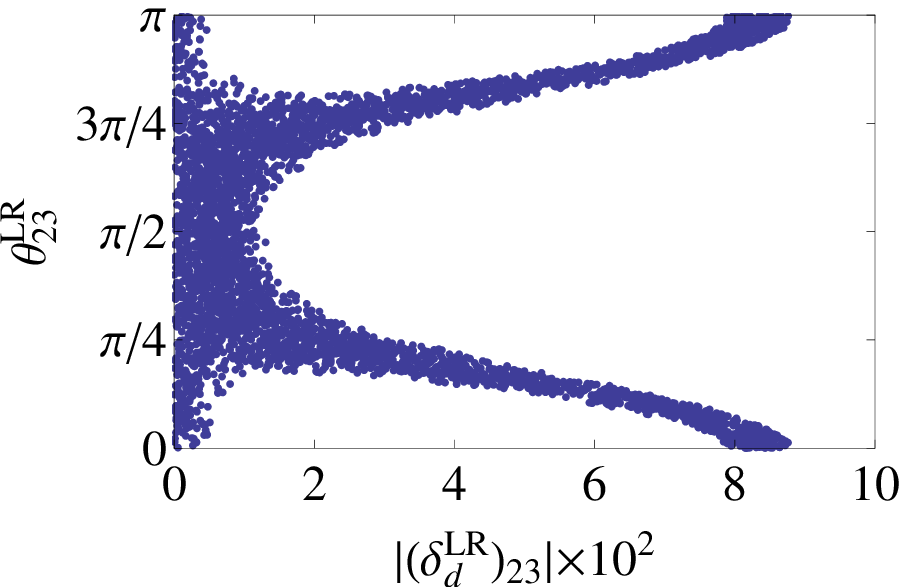}
\end{minipage}
\hspace{1cm}
\begin{minipage}[]{0.45\linewidth}
\hspace{4cm}(b)

\includegraphics[width=7.5cm]{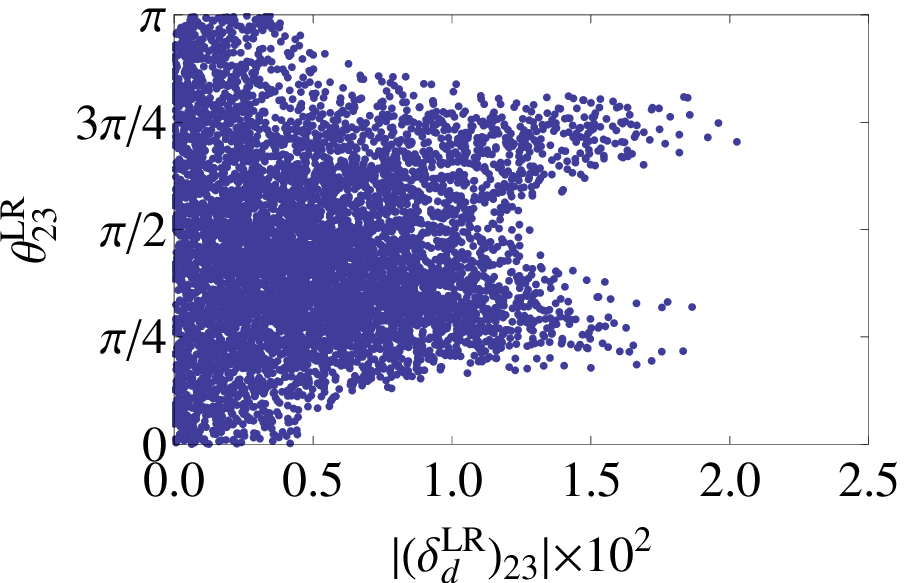}
\end{minipage}
\caption{The predicted region of $(\delta _d^{LR})_{23}$. 
In both figures (a) and (b), 
the horizontal and vertical axes denote the absolute value and
the phase of $(\delta _d^{LR})_{23}$, respectively. 
In the figure (a), 
the only experimental constraint of BR$(b\to s\gamma )$ 
is taken account. 
In the figure (b), 
the experimental constraints of BR$(b\to s\gamma )$, $S_{\phi K_S}$, 
and $S_{\eta 'K^0}$ are taken account.}
\label{fig:MIphase23}
\end{figure}

At first, in Fig.~\ref{fig:MIphase23} (a), 
we show the allowed region 
 in the plane of the absolute value $|(\delta _d^{LR})_{23}|$ and 
the phase $\theta_{23}^{LR}$ ,
where the only experimental constraint of $\text{BR}(b\to s\gamma )$ is put. 
The magnitude of the MI parameter $(\delta _d^{LR})_{23}$ is allowed as 
$|(\delta _d^{LR})_{23}|\lesssim 9\times 10^{-2}$.
We note that the SUSY contribution to $\text{BR}(b\to s\gamma )$  becomes superior compared with the SM 
in the region of $|(\delta _d^{LR})_{23}|\gtrsim 4\times 10^{-2}$. 
It is also noted that  the $(\delta_d^{LR})_{23}$ is almost real around
 the upper bound $9\times 10^{-2}$, that is at $2\theta _{23}^{LR}=0$ or $2\pi $. 
The experimental constraints of $S_{J/\psi K_S}$, $S_{\phi K_S}$, 
and $S_{\eta 'K^0}$ give the severe cut as seen 
in Fig.~\ref{fig:MIphase23} (b), where these experimental data
 are put in addition to $\text{BR}(b\to s\gamma )$.
In this figure, any value of the phase is allowed in 
$|(\delta _d^{LR})_{23}|\lesssim 5\times 10^{-3}$. 
On the other hand, the larger region of $|(\delta _d^{LR})_{23}|$ is allowed 
until $2\times 10^{-2}$ around the specific 
$\theta _{23}^{LR}$, $\pi /4$ and $3\pi /4$.\footnote{There still remains a very small allowed region 
around $|(\delta _d^{LR})_{23}|=9\times 10^{-2}$, where $(\delta _d^{LR})_{23}$ is almost real, 
since this region cannot be excluded by the time dependent CP asymmetries.
In our work, we omit this region hereafter. This region is uninteresting 
because the SUSY contribution is much larger than the SM one in $b\to s \gamma$.} 
The obtained  bound $|(\delta _d^{LR})_{23}|\lesssim 2\times 10^{-2}$
depends on the gluino and the squark masses. If they increase,
 the upper bound is rescaled approximately as
 $|(\delta _d^{LR})_{23}|\times m_{\tilde q}/(1.5~{\rm TeV})$.

\begin{figure}[h!]
\begin{minipage}[]{0.45\linewidth}
\vspace{3mm}
\includegraphics[width=7.5cm]{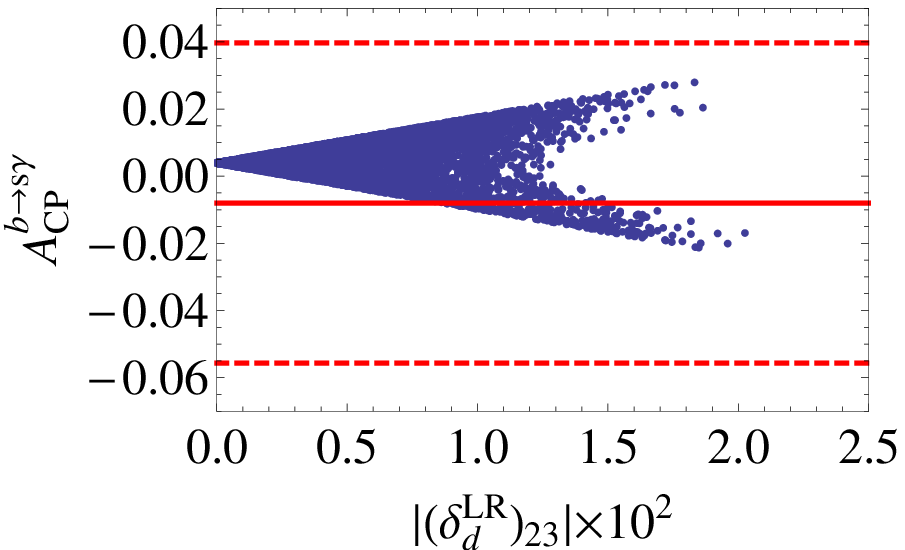}
\caption{The predicted direct CP asymmetry $A_{\text{CP}}^{b\to s\gamma }$ of $b\to s\gamma $ 
versus $|(\delta _d^{LR})_{23}|$. 
The red solid and two red dotted lines denote the best fit value, upper and lower bounds 
of the experimental data with $90\%$ C.L., respectively.}
\label{fig:MIdirectbs}
\end{minipage}
\hspace{1cm}
\begin{minipage}[]{0.45\linewidth}
\includegraphics[width=7.5cm]{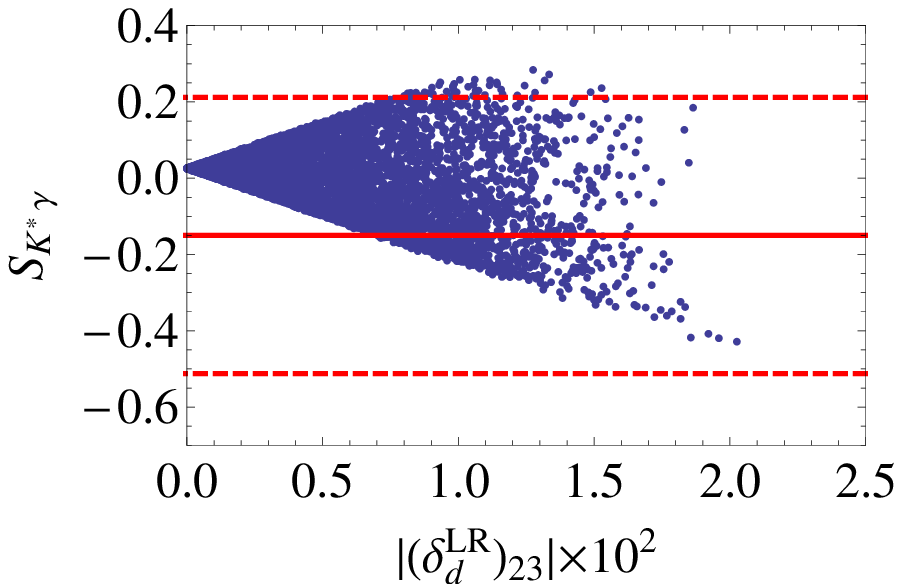}
\caption{The predicted CP asymmetry, $S_{K^* \gamma }$ of $B^0\to K^* \gamma $ 
versus $|(\delta _d^{LR})_{23}|$, where the red solid and two red dotted lines denote 
the best fit value, upper and lower bounds 
of the experimental data with $90\%$ C.L., respectively.}
\label{fig:MISKstargamma}
\end{minipage}
\end{figure}

By using this allowed region of $(\delta _d^{LR})_{23}$, 
we predict $A_{\text{CP}}^{b\to s\gamma }$, $S_{K^* \gamma }$, $S_{\phi \phi }$, 
and $S_{\phi \eta '}$. 
In Fig.~\ref{fig:MIdirectbs}, we show the predicted direct CP asymmetry 
$A_{\text{CP}}^{b\to s\gamma }$ versus $|(\delta _d^{LR})_{23}|$. 
Here the value at $|(\delta _d^{LR})_{23}|=0$ is the SM one, 
$A_{\text{CP}}^{b\to s\gamma }({\rm SM})\simeq 4\times 10^{-3}$~\cite{Kagan:1998bh}. 
We predict $-3\times 10^{-2}\lesssim A_{\text{CP}}^{b\to s\gamma }\lesssim 3\times 10^{-2}$ 
owing to the squark flavor mixing. 
Recent experimental data is still consistent with our prediction due to the 
large error as seen in $A_{\text{CP}}^{b\to s\gamma }(\text{exp}) = -0.008 \pm 0.029$~\cite{PDG}. 
The precise data will give us an additional constraint of the MI parameters in the future. 
 
In Fig.~\ref{fig:MISKstargamma}, we show the predicted CP asymmetry, $S_{K^* \gamma }$. 
The  predicted value in the SM is
 $S_{K^* \gamma }(\text{SM})\simeq (2m_s/m_b)\sin \phi _d \simeq  4\times 10^{-2}$~\cite{Atwood:1997zr}, 
while the experimental result is $S_{K^* \gamma }(\text{exp})= -0.15 \pm 0.22$~\cite{PDG}. 
Our prediction is $-0.4\lesssim S_{K^*\gamma }\lesssim 0.2$, 
which is still consistent with the experimental data. 
We also expect the precise data in the near future to test our prediction. 

\begin{figure}[h!]
\begin{minipage}[]{0.45\linewidth}
\vspace{3mm}
\includegraphics[width=7.5cm]{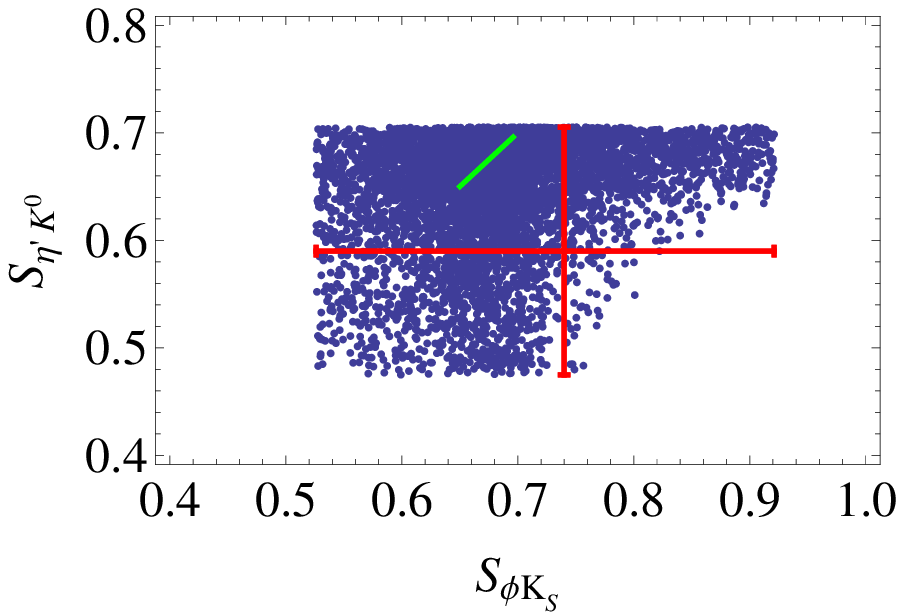}
\caption{The allowed region of the time dependent CP asymmetries 
on the $S_{\phi K_S}$--$S_{\eta 'K^0}$ plane. 
The SM prediction $S_{J/\psi K_S}=S_{\phi K_S}=S_{\eta 'K^0}$ 
is plotted by the green slant line. 
The experimental data with error bar is plotted by the red solid lines at $90\%$ C.L..}
\label{fig:SphiKSetapK}
\end{minipage}
\hspace{1cm}
\begin{minipage}[]{0.45\linewidth}
\vspace{-3mm}
\includegraphics[width=7.5cm]{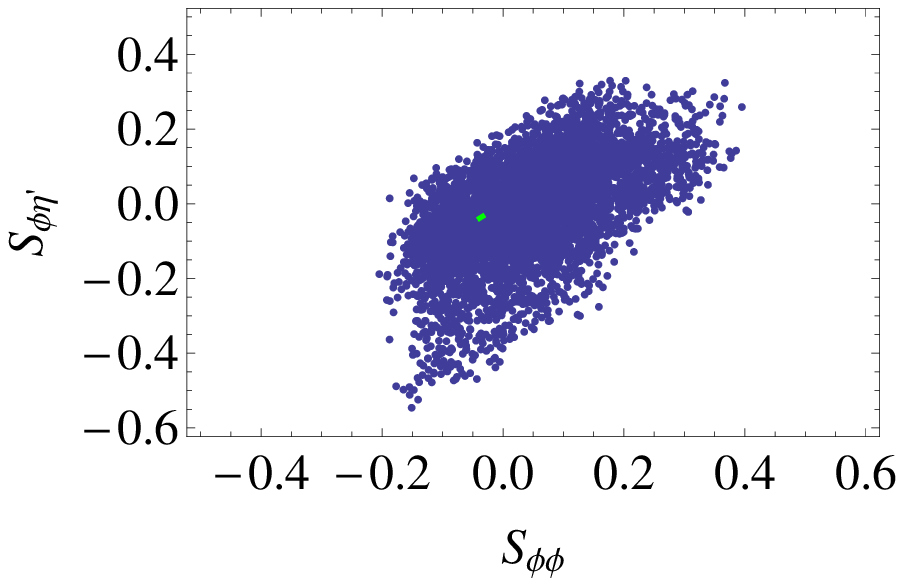}
\caption{The predicted time dependent CP asymmetries on the $S_{\phi \phi }$--$S_{\phi \eta '}$ plane. 
The small green line denotes the SM prediction from the experimental data of $S_{J/\psi \phi }$.}
\label{fig:SphiphiSphietap}
\end{minipage}
\end{figure}

Although the experimental data of the time dependent CP asymmetries 
$S_{\phi K_S}$ and $S_{\eta 'K^0}$ are taken as  the input in 
our analysis, these calculated values do not always cover 
all experimental allowed regions
due to the constraint from $\text{BR}(b\to s\gamma )$.
Those allowed regions are shown in  Fig.~\ref{fig:SphiKSetapK}. 
The SM prediction is $S_{J/\psi K_S}\text{(SM)}=S_{\phi K_S}\text{(SM)}=S_{\eta 'K^0}\text{(SM)}$, 
while the present data of these time dependent CP asymmetries are given in Eq.~(\ref{Sfdata}). 
The region of the right-down corner in the figure is excluded. 
It is testable in the future experiments. 
 
In Fig.~\ref{fig:SphiphiSphietap}, we predict the time dependent CP asymmetries 
$S_{\phi \phi }$ and $S_{\phi \eta '}$. 
These CP asymmetries must be equal to $S_{J/\psi \phi }$ in the SM. 
We use the experimental result of $S_{J/\psi \phi }$ for 
the phase $\phi _s$, which is given in Eq.~(\ref{phasedata}), in our calculations. 
We denote the small green line as the SM value 
$S_{J/\psi \phi }(\text{SM})=-0.0363^{+0.0016}_{-0.0015}$~\cite{Charles:2011va} in the figure. 
In conclusion, we predict $-0.2\lesssim S_{\phi \phi }\lesssim 0.4$ 
and $-0.5\lesssim S_{\phi \eta '}\lesssim 0.4$, respectively. 
Since the phase $\phi _s$ has still large experimental error bar, 
our prediction will be improved if the precise experimental data of 
$S_{J/\psi \phi }$ will be given in the near future at LHCb. 
Since the time dependent CP asymmetry $S_{\phi \phi }$ will be measured at LHCb, 
our prediction will be tested soon. 


\newpage 


\section{The $b\to d$ transition}
\label{sec:bdtransitions}
In this section, we discuss the $b\to d$ transition as the same way in the $b\to s$ one. 
The SUSY contribution is given in terms of the MI parameters 
$(\delta _d^{LL})_{13}$, $(\delta _d^{RR})_{13}$, $(\delta _d^{LR})_{13}$, and 
$(\delta _d^{RL})_{13}$. 
The typical  $b\to d$ transition  is the $b\to d\gamma $ decay. 
The experimental data of its branching ratio gives the constraint for these MI parameters. 
By using these MI parameters, 
we calculate SUSY contributions to the direct CP violation of 
the $b\to d\gamma $ decay and the time dependent CP asymmetry 
in the $B^0\to \rho \gamma $ decay. 
We also predict the time dependent CP asymmetry of 
the $B^0\to K^0\bar K^0$ decay. 

In order to constrain the MI parameters, we input the experimental data  
of the branching ratio of $b\to d\gamma $~\cite{delAmoSanchez:2010ae,Crivellin:2011ba}, 
\begin{equation}
\text{BR}(b\to d\gamma )({\rm exp})=(1.41\pm 0.57)\times 10^{-5},
\label{BRbdgamma}
\end{equation}
on the other hand, the SM has predicted~\cite{Crivellin:2011ba}
\begin{equation}
\text{BR}(b\to d\gamma )({\rm SM})=(1.54_{-0.31}^{+0.26})\times 10^{-5}.
\end{equation}

Next we present the formulations of the time dependent CP asymmetries 
and direct CP violation including SUSY contributions. 
The branching ratio and direct CP violation in the $b\to d\gamma $ decay 
are given in Eqs.~(\ref{Brbqgamma}) and (\ref{directbqgamma}), respectively. 
The time dependent CP asymmetry $S_{\rho \gamma }$ in the 
$B^0 \to \rho \gamma $ decay is an important observable to search for 
the new physics and given as 
\begin{equation}
S_{\rho \gamma }
=
\frac{2 {\rm Im}(-e^{-i\phi _d} {\tilde C}_{7\gamma }(m_b)/C_{7\gamma }(m_b))}
{|{\tilde C}_{7\gamma }(m_b)/C_{7\gamma }(m_b)|^2+1}.
\end{equation}
Since ${\tilde C}_{7\gamma }^\text{SM}(m_b)/C_{7\gamma }^\text{SM}(m_b)\propto m_d/m_b$ in the SM,
$S_{\rho \gamma }$ may be expected to be 
quite suppressed~\cite{Atwood:1997zr}. However, $S_{\rho \gamma }$ 
could be also enhanced owing to the gluino-squark mediated flavor changing
process.

The time dependent CP asymmetries $S_{K^0\bar K^0}$ and $C_{K^0\bar K^0}$ in 
the $B^0\to K^0\bar K^0$ decay 
are also interesting ones to search for the new physics 
since there is no tree process of the SM in the $B^0\to K^0\bar K^0$ decay 
~\cite{Giri:2004af,Fleischer:2004vu}. 
These CP asymmetries are given in Eq.~(\ref{sf}) as 
\begin{equation}
S_{K^0\bar K^0}=\frac{2\text{Im}\lambda _{K^0\bar K^0}}
{1+|\lambda _{K^0\bar K^0}|^2}~,
\qquad 
C_{K^0\bar K^0}=\frac{1-|\lambda _{K^0\bar K^0}|^2}{1+|\lambda _{K^0\bar K^0}|^2}~,
\end{equation}
where 
\begin{equation}
\lambda _{K^0\bar K^0}=\frac{q}{p} \bar \rho ~, \qquad 
\frac{q}{p}\simeq \sqrt{\frac{M_{12}^{d*}}{M_{12}^{d}}}, \qquad 
\bar \rho \equiv 
\frac{\bar A(\bar B^0 \to K^0 \bar K^0)}{A(B^0\to K^0 \bar K^0)}.
\end{equation}
The amplitude $\bar A(\bar B^0\to K^0\bar K^0)$ 
is given in Ref.~\cite{Giri:2004af}, 
\footnote{The $\bar A(\bar B^0\to K^0\bar K^0)$ 
amplitude is explicitly presented in Refs.~\cite{Giri:2004af,Muta:2000ti}. 
In our calculation, we neglect $C_i$ $(i=8-10)$ since these Wilson coefficients are too small 
to contribute to the amplitude of $\bar B^0\to K^0\bar K^0$ in our model.} 
in which the QCD corrections are important for the hadronic matrix elements~\cite{Muta:2000ti}, as 
\begin{equation}
\bar A(\bar B^0\to K^0\bar K^0)
\simeq 
\frac{4G_F}{\sqrt{2}}\sum _{q=u,c}V_{qb}V_{qd}^*
\left [a_4^q(m_b)+r_\chi a_6^q(m_b)\right ]X.
\end{equation}
Here $X$ is the factorized matrix element (See Ref.~\cite{Giri:2004af}.) as 
\begin{equation}
X=-if_KF_0(m_K^2)(m_B^2-m_K^2),
\end{equation}
where $f_K$ and $F_0(m_K^2)$ denote the decay coupling constant of the 
$K$ meson and the form factor, respectively, 
and $r_\chi=2m_K^2/((m_b-m_s)(m_s+m_d))$ denotes the chiral enhancement factor. 
The coefficients $a_i^q$'s are given as~\cite{Giri:2004af,Muta:2000ti} 
\begin{align}
a_4^q(m_b)&=(C_4-\tilde C_4)+\frac{(C_3-\tilde C_3)}{N_c}+\frac{\alpha _s(m_b)}{4\pi }\frac{C_F}{N_c}
\Bigg [(C_3-\tilde C_3)\left [F_K+G_K(s_d)+G_K(s_b)\right ] \nonumber \\
&\hspace{1cm}+C_2G_K(s_q)+\left [ (C_4-\tilde C_4)+(C_6-\tilde C_6)\right ] 
\sum _{f=u}^bG_K(s_f)+(C_{8G}-\tilde C_{8G})G_{K,g}\Bigg ], \nonumber \\
a_6^q(m_b)&=(C_6-\tilde C_6)+\frac{(C_5-\tilde C_5)}{N_c}+\frac{\alpha _s(m_b)}{4\pi }\frac{C_F}{N_c}
\Bigg [(C_3-\tilde C_3)\left [G_K'(s_d)+G_K'(s_b)\right ] \nonumber \\
&\hspace{1cm}+C_2G_K'(s_q)+\left [(C_4-\tilde C_4)+(C_6-\tilde C_6)\right ]
\sum _{f=u}^bG_K'(s_f)+(C_{8G}-\tilde C_{8G})G_{K,g}'\Bigg ],
\label{coefficients-BKK}
\end{align}
where $q$ takes $u$ and $c$ quarks, $C_F=(N_c^2-1)/(2N_c)$, 
and the loop functions $F_K$, $G_K$, $G_{K,g}$, $G_K'$, 
and $G_{K,g}'$ are given in Refs.~\cite{Giri:2004af,Muta:2000ti}. 
The internal quark mass in the penguin diagrams 
enters as $s_f=m_f^2/m_b^2$.\footnote{The $C_i^{\tilde g}~(i=3-6,8G)$ 
in Eq.~(\ref{coefficients-BKK}) should be taken as the replacement
 $C_i^{\tilde g}\rightarrow [(V_{tb}V_{td}^*)/(V_{qb}V_{qd}^*) ]C_i^{\tilde q}$ 
in Eq.~(\ref{Coeff}).} The minus sign in front of $\tilde C_i~(i=3-6,8G)$ comes from the 
parity of the final state as discussed in the previous section. 

By using above formulations, we estimate the SUSY contributions in the $b\to d$ 
transition. 
In our calculations, we take $\mu \tan \beta $ to be $1$~TeV and we set the MI parameters, 
$|(\delta _d^{LL})_{13}| = |(\delta _d^{LL})_{13}|\lesssim 10^{-2}$
from our previous works~\cite{Hayakawa:2012ua,Shimizu:2012ru}. 
We also assume that the magnitudes of the MI parameters $(\delta _d^{LR})_{13}$ and 
$(\delta _d^{RL})_{13}$ are same but each phase is different. 
Thus, we  parameterize the MI parameters as follows: 
\begin{equation}
(\delta _d^{LR})_{13}=|(\delta _d^{LR})_{13}|e^{2i\theta _{13}^{LR}},\quad 
(\delta _d^{RL})_{13}=|(\delta _d^{LR})_{13}|e^{2i\theta _{13}^{RL}}.
\end{equation}

\begin{figure}[h!]
\begin{minipage}[]{0.45\linewidth}
\includegraphics[width=7.5cm]{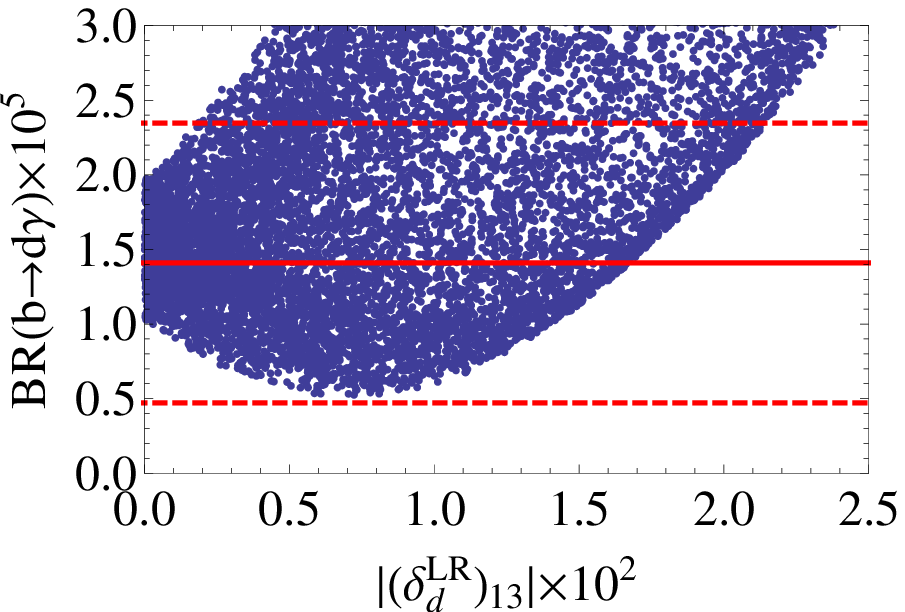}
\caption{The predicted region on the $|(\delta _d^{LR})_{13}|$--$\text{BR}(b\to d\gamma )$ plane. 
The red solid and two red dotted lines denote the best fit value, upper and lower bounds 
of the experimental data with $90\% $~C.L., respectively.}
\label{fig:MIBRbdgamma}
\end{minipage}
\hspace{1cm}
\begin{minipage}[]{0.45\linewidth}
\vspace{-1.7cm}
\includegraphics[width=7.5cm]{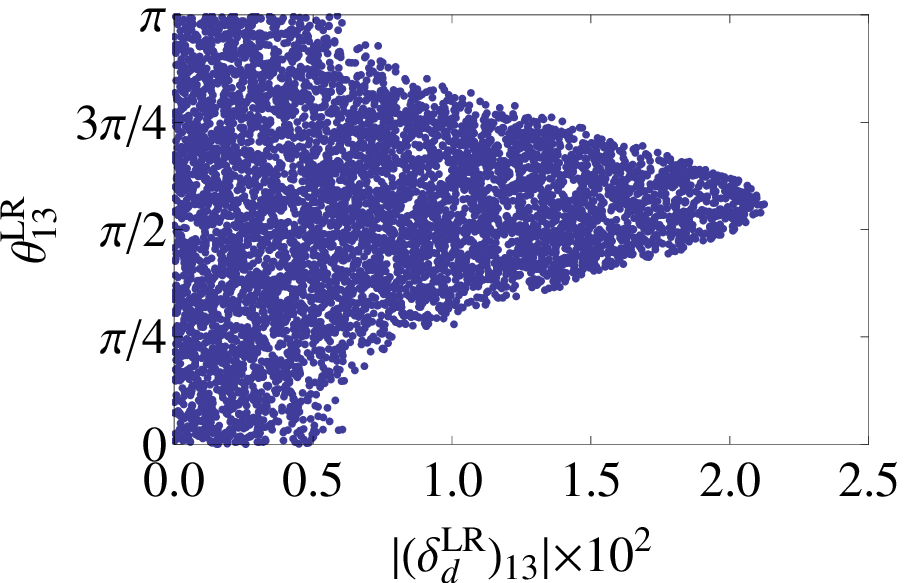}
\caption{The predicted region on the $(\delta _d^{LR})_{13}$--$\theta _{13}^{LR}$ plane. 
The experimental constraint of BR$(b \to d \gamma)$ is taken account.}
\label{fig:MIphase13} 
\end{minipage}
\end{figure}

Let us discuss the numerical analysis. In our calculations, 
we use the squark mass and the gluino mass as given in  Eq.~(\ref{SUSYmass}). 
The present experimental data of BR$(b\to d\gamma )$ in Eq.~(\ref{BRbdgamma}) gives a 
constraint for the MI parameters as seen in Fig.~\ref{fig:MIBRbdgamma}. 
The SM contribution is larger than the SUSY one 
until $|(\delta _d^{LR})_{13}|\simeq 7\times 10^{-3}$, while
the SUSY contribution dominates the $b\to d\gamma $  decay
in the region of $|(\delta _d^{LR})_{13}|\gtrsim 7\times 10^{-3}$.
It is remarked that there is a lower bound of the branching ratio 
around $5\times 10^{-6}$. 

In Fig.~\ref{fig:MIphase13}, we show the allowed region of $(\delta _d^{LR})_{13}$ 
within $90\% $ C.L. of BR$(b\to d\gamma )$. It is found that any value of the phase is allowed in $|(\delta _d^{LR})_{13}|\lesssim 5\times 10^{-3}$.
The upper bound of the MI parameter is at 
$|(\delta _d^{LR})_{13}|\simeq 2\times 10^{-2}$ around the specific $\theta _{13}^{LR}$, 
 $\pi /2$. 

\begin{figure}[h!]
\begin{minipage}[]{0.45\linewidth}
\includegraphics[width=7.5cm]{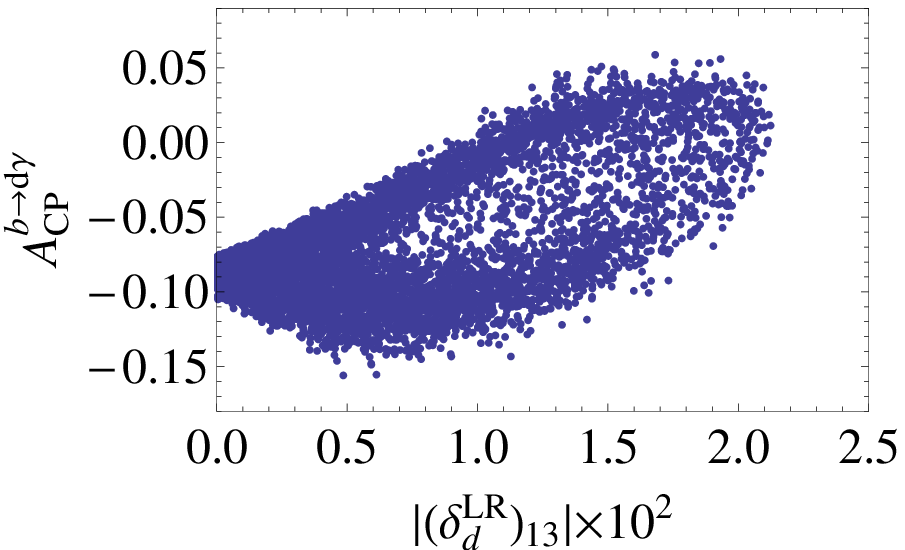}
\caption{The predicted direct CP asymmetry $A_{\text{CP}}^{b \to d\gamma }$ 
versus $|(\delta _d^{LR})_{13}|$.}
\label{fig:MIdirectbd}
\end{minipage}
\hspace{1cm}
\begin{minipage}[]{0.45\linewidth}
\vspace{-3mm}
\includegraphics[width=7.5cm]{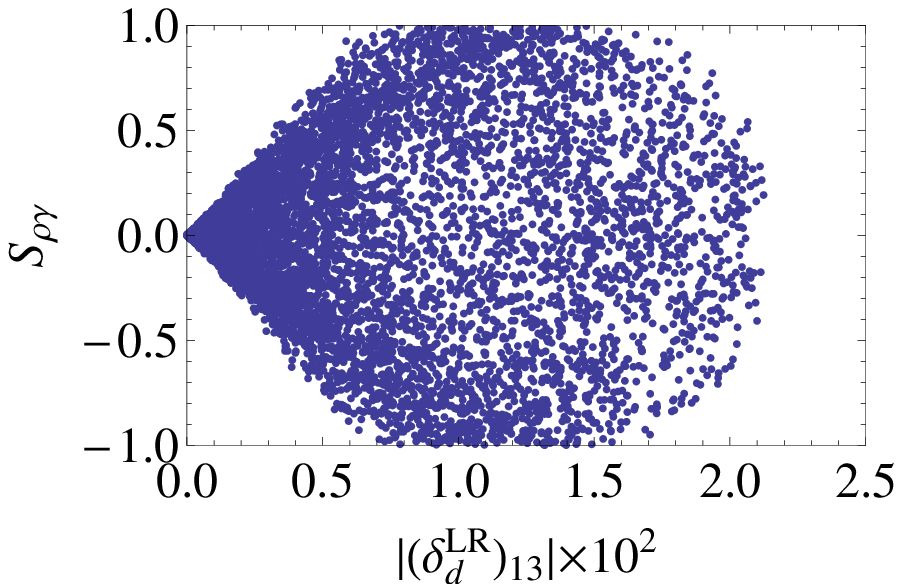}
\caption{The predicted time dependent CP asymmetry 
$S_{\rho \gamma }$ versus $|(\delta _d^{LR})_{13}|$.}
\label{fig:MISrhogamma}
\end{minipage}
\end{figure}

By using this allowed region of $(\delta _d^{LR})_{13}$, we can predict 
the direct CP asymmetry $A_\text{CP}^{b\to d\gamma }$ and time dependent CP asymmetries 
$S_{\rho \gamma }$, $S_{K^0\bar K^0}$, and $C_{K^0\bar K^0}$. 
In Fig.~\ref{fig:MIdirectbd}, we show the predicted direct CP asymmetry 
$A_\text{CP}^{b\to d\gamma }$ versus $|(\delta _d^{LR})_{13}|$. 
Here the value at $|(\delta _d^{LR})_{13}|=0$ is 
the SM one, $A_\text{CP}^{b\to d\gamma }(\text{SM})\simeq -0.09$. 
 Our prediction is 
$-0.16\lesssim A_\text{CP}^{b\to d\gamma }\lesssim 0.06$.
If $A_\text{CP}^{b\to d\gamma }$ is measured in the future, 
we obtain an additional  constraint of the MI parameters. 

In Fig.~\ref{fig:MISrhogamma}, we show the prediction of $S_{\rho \gamma }$ 
depending on $|(\delta _d^{LR})_{13}|$. 
The SM prediction is $S_{\rho \gamma }(\text{SM})\simeq (2m_d/m_b)
\sin \phi _d\simeq 2.0\times 10^{-3}$~\cite{Atwood:1997zr}, 
while the experimental data is 
$S_{\rho \gamma }(\text{exp})=-0.8\pm0.7$~\cite{PDG}. 
In our prediction, the $S_{\rho \gamma }$ reaches $\pm 1$ at 
$|(\delta _d^{LR})_{13}|\gtrsim 7\times 10^{-3}$. 
Therefore,
 the $S_{\rho \gamma }$ is expected to be much larger than 
the SM prediction in the case of  $|(\delta _d^{LR})_{13}|=  {\cal O}(10^{-3})$. 
We expect the precise data to test our prediction in the future. 

\begin{figure}[h!]
\begin{minipage}[]{0.45\linewidth}
\vspace{-5mm}
\includegraphics[width=7.5cm]{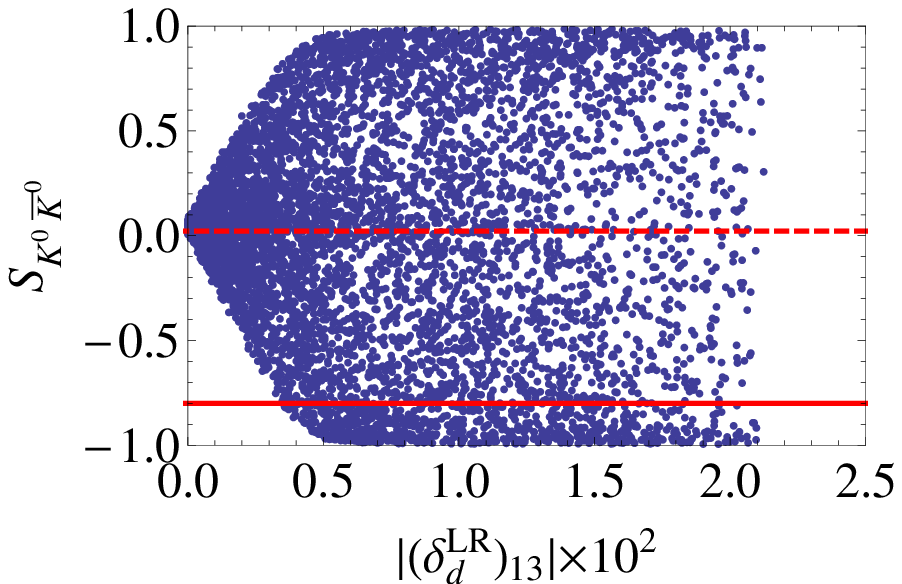}
\caption{The predicted time dependent CP asymmetry $S_{K^0\bar K^0}$ 
versus $|(\delta _d^{LR})_{13}|$. 
The red solid and red dotted lines denote the best fit value 
and the experimental data with $90\% $~C.L., respectively.}
\label{fig:MISKK}
\end{minipage}
\hspace{1cm}
\begin{minipage}[]{0.45\linewidth}
\includegraphics[width=7.5cm]{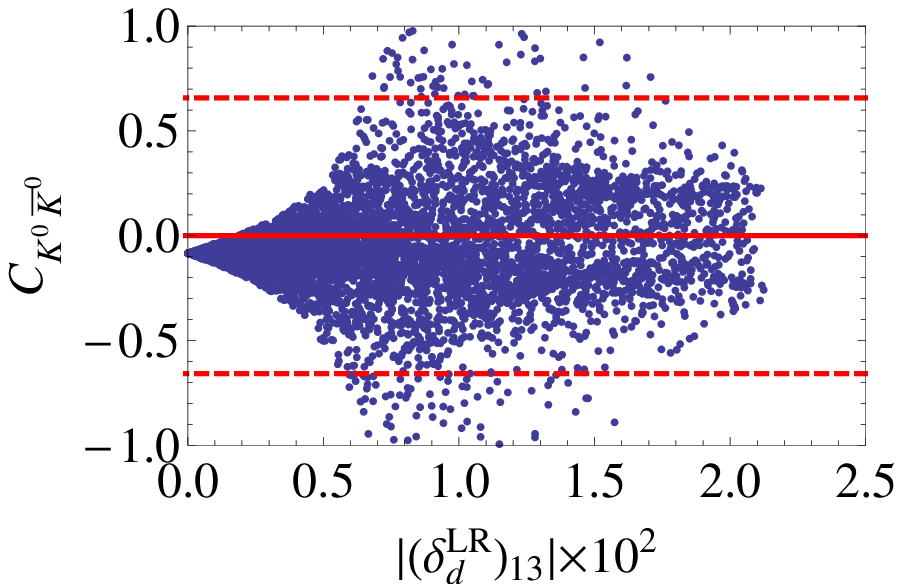}
\caption{The predicted time dependent CP asymmetry $C_{K^0\bar K^0}$ 
versus $|(\delta _d^{LR})_{13}|$. 
The red solid and two red dotted lines denote the best fit value, upper and lower bounds 
of the experimental data with $90\% $~C.L., respectively.}
\label{fig:MICKK}
\end{minipage}
\end{figure}

In Figs.~\ref{fig:MISKK} and \ref{fig:MICKK}, 
we show the predictions of the time dependent CP asymmetries 
$S_{K^0\bar K^0}$ and $C_{K^0\bar K^0}$ depending on $|(\delta _d^{LR})_{13}|$,
respectively. 
In the SM, one predicts $0.02 \le S_{K^0\bar K^0}(\text{SM})\le0.13$ 
and $-0.17 \le C_{K^0\bar K^0}(\text{SM})\le -0.15$~\cite{Giri:2004af}, 
while the experimental data are given as $S_{K^0\bar K^0}(\text{exp})=-0.8\pm 0.5$ 
and $C_{K^0\bar K^0}(\text{exp})=0.0\pm 0.4$~\cite{PDG}, respectively. 
The present experimental bounds do not give any additional constraints to 
Fig.~\ref{fig:MIphase13}. 
However, more precise experimental data provide intensive constraints for MI parameters. 




\begin{table}[t]
\begin{center}
\begin{tabular}{|c||c|c|c|}
\hline
 & Exp. & SM & our prediction \\
\hline 
\hline
BR$(b \to s \gamma)$ & $(3.53\pm 0.24)\times 10^{-4}$ \cite{PDG}& $(3.15 \pm 0.23)\times 10^{-4} $~\cite{Misiak:2006zs} & constraint \\
BR$(b \to d \gamma )$ & $(1.41\pm 0.57)\times 10^{-5}$~\cite{delAmoSanchez:2010ae,Crivellin:2011ba} & $(1.54_{-0.31}^{+0.26})\times 10^{-5}$~\cite{Crivellin:2011ba} & constraint \\
\hline
$A_\text{CP}^{b \to s \gamma}$ & $-0.008 \pm 0.029$ \cite{PDG} & $4\times 10^{-3}$~\cite{Kagan:1998bh} & $-0.03\sim 0.03$ \\
$A_\text{CP}^{b \to d \gamma}$ & ------ & $-0.09$ & $-0.16\sim 0.06$ \\
\hline
$S_{J/\psi K_S}$ & $0.679\pm0.020$~\cite{Amhis:2012bh} & input & constraint \\
$S_{\phi K_S}$ & $0.74^{+0.11}_{-0.13}$~\cite{Amhis:2012bh} & $=S_{J/\psi K_S}$ & constraint \\
$S_{\eta' K^0}$ & $0.59\pm{0.07}$~\cite{Amhis:2012bh}         & $=S_{J/\psi K_S}$   & constraint \\
$\phi_s (S_{J/\psi \phi }=\sin \phi _s)$ & $-0.004\pm 0.166\pm 0.054$~\cite{ICHEP2012} 
& $-0.0363^{+0.0016}_{-0.0015}$ \cite{Charles:2011va} & constraint \\
$S_{\phi \phi }$ & ------ & $=S_{J/\psi \phi }$ & $-0.2\sim 0.4$\\
$S_{\phi \eta '}$ & ------ & $=S_{J/\psi \phi }$ & $-0.5\sim 0.4$ \\
$S_{K^* \gamma}$ & $-0.15 \pm 0.22$~\cite{PDG} & $0.04$~\cite{Atwood:1997zr} & $-0.4\sim 0.2$ \\
$S_{\rho \gamma }$ & $-0.8\pm 0.7$~\cite{PDG} & $0.002$~\cite{Atwood:1997zr} & $-1\sim 1$ \\
$S_{K^0\bar K^0}$ & $-0.8\pm 0.5$~\cite{PDG} & $0.02\sim 0.13$~\cite{Giri:2004af} 
& $-1\sim 1$ \\
$C_{K^0\bar K^0}$ & $-0.0\pm 0.4$~\cite{PDG} & $-0.17\sim -0.15$~\cite{Giri:2004af} & $-1\sim 1$ \\
\hline
\end{tabular}
\end{center}
\caption{Summary of the SM predictions, experimental values, and our predictions.}
\label{tab:summary}
\end{table}

\section{Summary}

We have discussed the contribution of the gluino-squark mediated flavor changing
process to the CP violation 
in $b\to s$ and $b\to d$ transitions taking account of recent experimental data. 
We have presented the allowed region of the MI parameters 
$(\delta _d^{LR})_{23}$ and $(\delta _d^{LR})_{13}$, which are constrained 
by the branching ratios of  $b\to s\gamma$ and $b\to d\gamma$ decays. 
In addition, the time dependent CP asymmetries of 
$B^0\to  J/\psi K_S$, $B^0\to  \phi K_S$, and $B^0\to  \eta ' K^0$ decays
 severely
restrict the allowed region of the MI parameter, $(\delta _d^{LR})_{23}$. 
These MI parameters $(\delta _d^{LR})_{23}$ and $(\delta _d^{LR})_{13}$ 
are still allowed up to $2\times 10^{-2}$ 
for the squark and gluino masses of $1.5$~TeV. 
If $m_{\tilde q}\simeq m_{\tilde g}$ increase, the bound of $(\delta _d^{LR})_{k3}~(k=2,1)$ 
is approximately rescaled as $(\delta _d^{LR})_{k3}\times m_{\tilde q}/(1.5~\text{TeV})$. 

By using these constraints, we predict the CP asymmetries of 
$B_s\to \phi \phi$, $B_s\to \eta '\phi $, and $B^0\to K^0 \bar K^0$ decays, 
as well as the CP asymmetries in $b\to s\gamma $ and $b\to d\gamma $ decays. 
We have summarized our results in Table~\ref{tab:summary}. 
It is remarked that the CP violation of the  $B_s\to  \phi \phi$ decay
 is expected to be large owing to the squark flavor mixing.
  This prediction will be tested soon at LHCb.

\vspace{0.5 cm}
\noindent
{\bf Acknowledgment}

We thank S. Mishima  for useful discussions. 
We also thank A. Hayakawa and J. Kumagai for their help.
M.T. is  supported by JSPS Grand-in-Aid for Scientific Research,
21340055 and 24654062.



\end{document}